\documentclass[11pt,twoside]{article}
\usepackage{./asp2014}

\aspSuppressVolSlug
\resetcounters

\begin{document}

\title{ngVLA Studies of Classical Novae}
\author{Justin D. Linford$^{1,2}$, Laura Chomiuk,$^3$, and Michael P. Rupen$^4$
\affil{$^1$Department of Physics, The George Washington University, Washington, DC 20052, USA; \email{jlinford@gwu.edu}}
\affil{$^2$Astronomy, Physics, and Statistics Institute of Sciences (APSIS), 725 21st St NW, Washington, DC 20052, USA}
\affil{$^3$Center for Data Intensive and Time Domain Astronomy, Department of Physics and Astronomy, Michigan State University, 567 Wilson Road, East Lansing, MI 48824, USA; \email{chomiuk@pa.msu.edu}}
\affil{$^4$Herzberg Astronomy and Astrophysics, National Research Council, Penticton, BC, Canada; \email{michael.rupen@nrc-cnrc.gc.ca}}}

% This section is for ADS Processing.  There must be one line per author.
\paperauthor{Justin D. Linford}{jlinford@gwu.edu}{0000-0002-3873-5497}{The George Washington University}{Physics}{Washington}{DC}{20052}{USA}
\paperauthor{Laura Chomiuk}{chomiuk@pa.msu.edu}{0000-0002-8400-3705}{Michigan State University}{Physics and Astronomy}{East Lansing}{MI}{48824}{USA}
\paperauthor{Michael P. Rupen}{michael.rupen@nrc-cnrc.gc.ca}{}{Herzberg Astronomy and Astrophysics Research Centre}{}{Penticton}{BC}{}{Canada}

%Chapters do not have abstracts
%\begin{abstract}
%\end{abstract}

\section{Introduction}

A nova outburst results when sufficient mass accretes onto the surface of a white dwarf from a companion star, triggering a thermonuclear explosion.  In classical novae the bulk of the emission comes from the warm, expanding ejecta---$10^{-7}-10^{-3}$ M$_{\odot}$ expanding at a few thousand km/s.  The prevailing theories assume that the explosion occurs as a single, spherically symmetric ejection event and predict a simple relationship between the white dwarf mass, the accretion rate, and the ejecta mass and energetics of the explosion (e.g., Yaron et al. 2005).  However, observations with modern instruments indicate that nova eruptions are far from simple.  There is now evidence for multiple ejection events (e.g., Nelson et al. 2014), common envelopes (e.g., Chomiuk et al. 2014), non-spherical geometry (e.g., Slavin et al. 1995; Linford et al. 2015), and even jet-like structures in the ejecta (e.g., Rupen et al. 2008).  %The ENova collaboration combines radio, mm, optical, X-ray, and $\gamma$-ray observations and detailed theoretical modelling to study these most common major explosions in the universe.  Among our results so far are the direct demonstration of the importance of shocks in novae, including detection of $\gamma$-ray producing shocks in several sources (e.g., Chomiuk et al. 2014; Weston et al. 2016a,b), and the realization that multiple, long-lived outflows are much more common than previously assumed in these explosions (e.g., Nelson et al. 2014).  %Here we propose a new approach to characterize the early shocks in these complex explosions with the next generation Very Large Array.

One of the most surprising discoveries of the \emph{Fermi Gamma-ray Space Telescope} was that novae produce GeV $\gamma$-ray emission.  While MeV emission from novae was predicted (and yet never observed; Hernanz 2013), GeV emission requires a population of accelerated particles that was not expected to be present in nova explosions.  The first nova detected by \emph{Fermi}/LAT was V407 Cyg, which had a Mira giant companion (Abdo et al. 2010).  The nova was therefore embedded in the dense wind of its companion star, providing an ideal environment for producing strong shocks when the fast nova ejecta slammed into the slow wind material.  However, even novae with main sequence companions are capable of producing detectable $\gamma$-rays (e.g., Ackermann et al. 2014; Cheung et al. 2016), which is surprising given the negligible stellar winds leading to low density circumstellar material surrounding these binaries.  Recent results from studies at radio wavelengths demonstrate of the importance of shocks in novae, including detection of $\gamma$-ray producing shocks in several sources (e.g., Chomiuk et al. 2014; Weston et al. 2016a,b), and the realization that multiple, long-lived outflows are much more common than previously assumed in these explosions (e.g., Nelson et al. 2014).  Here we illustrate how the next generation Very Large Array will contribute to the study of classical novae, with an emphasis on characterizing the shocks that develop early in the evolution of these complex explosions.

\section{The Case of V959 Mon}
\label{sec:novae_v959mon}

\begin{figure}[t]
\begin{center}
\includegraphics[width=0.9\textwidth]{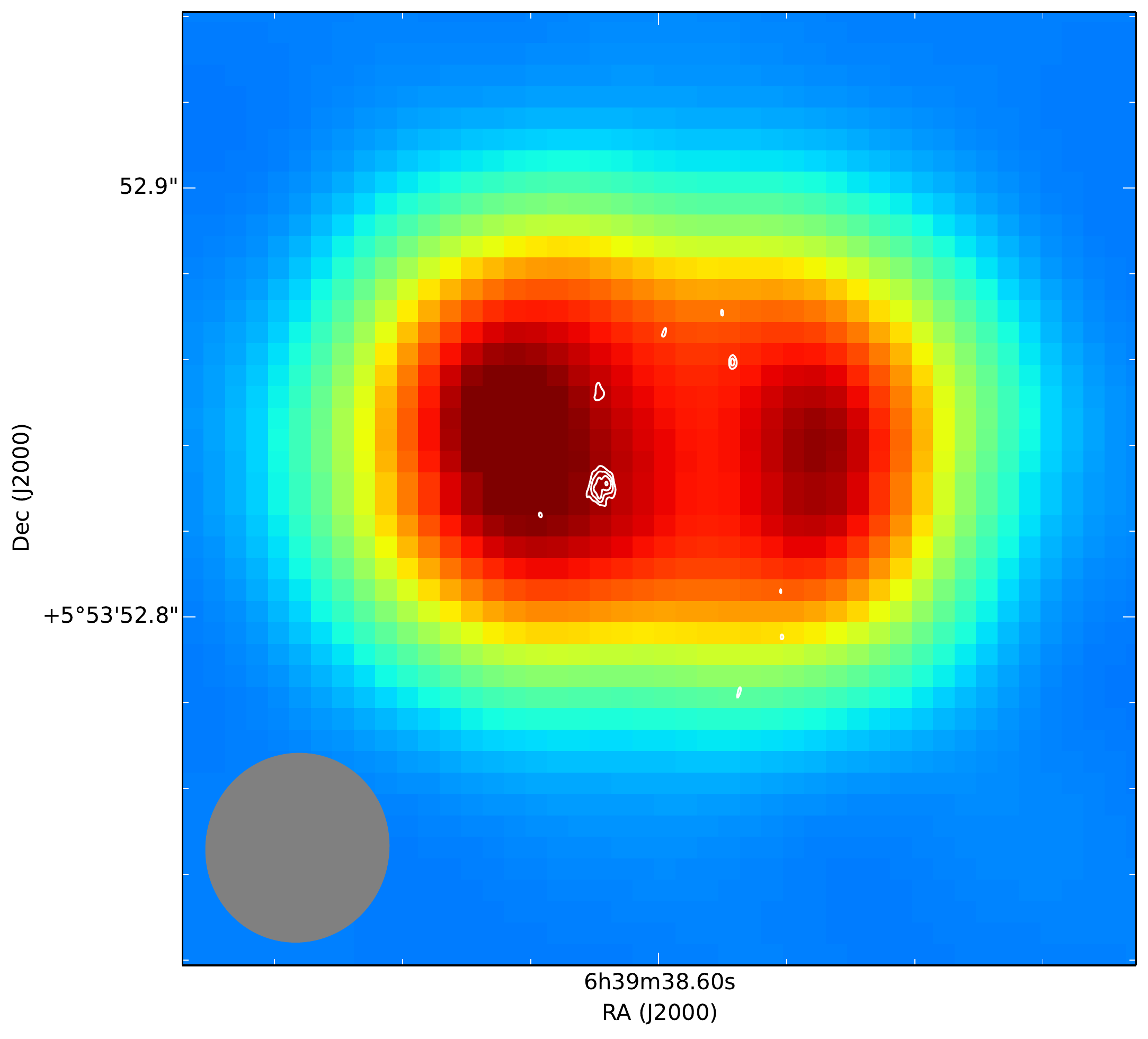}
\end{center}
\vspace{-0.6cm}
\caption{\textit{Imaging of the gamma-ray detected nova V959 Mon (2012), with compact regions of synchrotron emission (white contours) superimposed on the thermal ejecta (color scale). The synchrotron contours were imaged by the VLBA at 5 GHz 106 days after discovery, while the morphology of the thermal ejecta was revealed by the Jansky VLA at 36.5 GHz, 126 days after discovery.  The gray ellipse in the lower left corner represents the 36.5 GHz VLA restoring beam.  Adapted from Chomiuk et al. 2014.}}
\label{Mon_over}
\end{figure}

\begin{figure}[t]
\begin{center}
%\vspace{-0.5cm}
\includegraphics[width=0.9\textwidth]{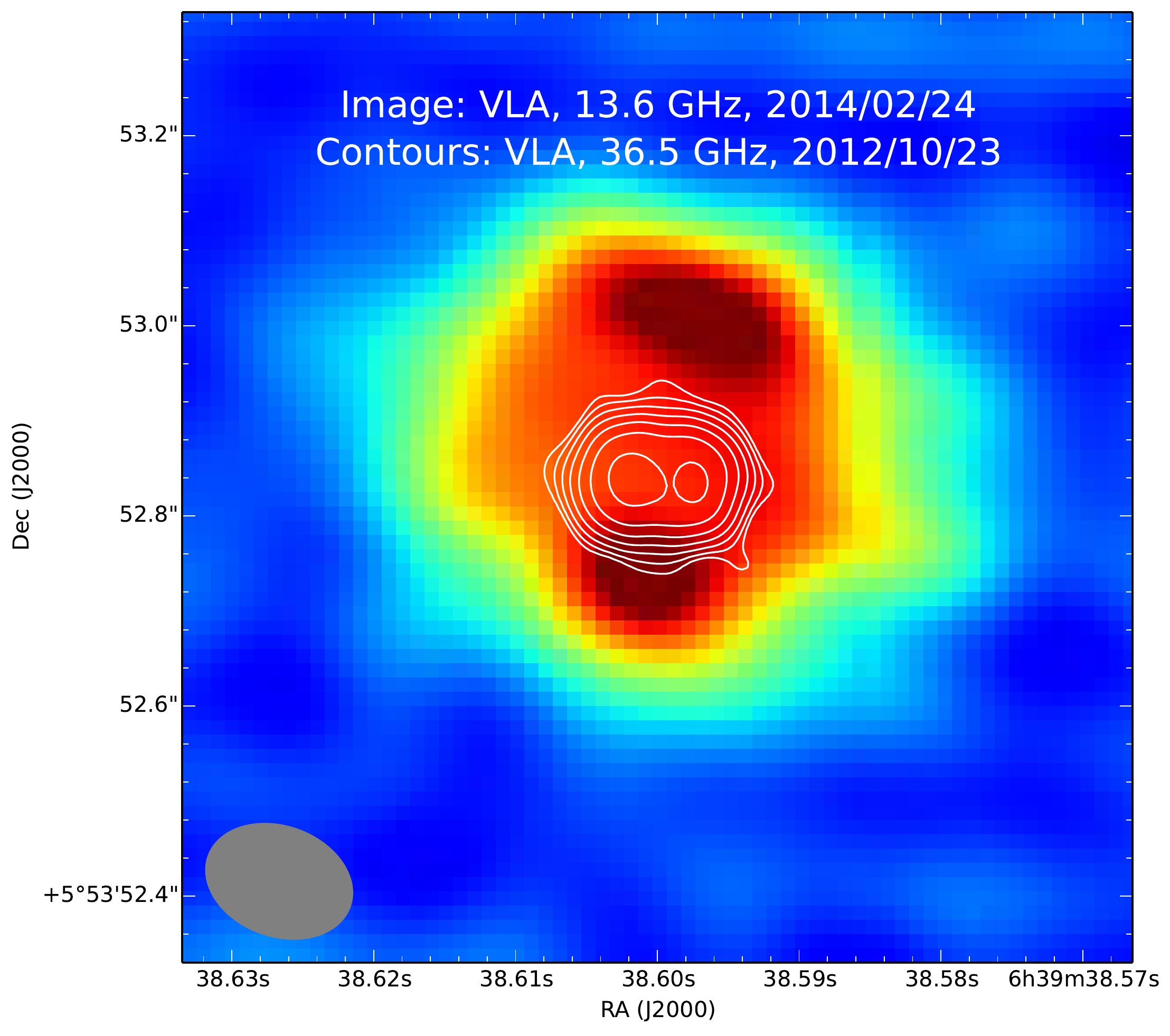}
\vspace{-0.3cm}
\caption{\textit{The evolution of V959 Mon over the two years following nova explosion, as imaged by the Jansky VLA which is sensitive to the thermal ejecta. The plane of the binary system is oriented in the N-S (up-down) direction.  \textbf{White Contours} show the 36.5 GHz image of V959 Mon 126 days after discovery, which is dominated by a fast bipolar outflow perpendicular to the orbital plane. \textbf{Color} shows the 13.6 GHz image of V959 Mon 615 days after discovery, which is dominated by the slowly expanding torus oriented along the orbital plane.  The Gray ellipse in the lower left represents the 13.6 GHz VLA restoring beam. Adapted from Chomiuk et al. 2014.}\vspace{-18pt}}
\label{Mon_2A}
\end{center}
\end{figure}

The 2012 nova V959~Mon was one of the first classical novae with a main sequence companion detected by \textit{Fermi}.  In fact, the nova was discovered by Fermi as it was too close to the Sun for optical observations.  Chomiuk et al. (2014) were able to directly image the shocked regions with the Very Long Baseline Array (VLBA) and European VLBI Network (EVN).  As shown in Figure~\ref{Mon_over}, the synchrotron-emitting compact regions did not align with the bipolar structure of the expanding thermal gas, as imaged by the Karl G. Jansky Very Large Array (VLA).  Instead, the combined VLBA, EVN, and VLA observations revealed that the compact components were moving on a path roughly 45$^{\circ}$ to the VLA structure.  High-resolution observations with the VLA separated by more than a year also revealed an apparent 90$^{\circ}$ flip in the orientation of the bipolar outflow (Figure~\ref{Mon_2A}).  Chomiuk et al. (2014) concluded that the nova ejecta had two distinct components: a slow flow in the equatorial plane of the binary with a velocity of $\sim480$ km s$^{-1}$, and a faster flow in the polar direction with a velocity of $\sim1200$ km s$^{-1}$.  Analysis of high-resolution VLA images of these two outflows led to a distance estimate of $1.4 \pm 0.4$ kpc (Linford et al. 2015).  The interaction between these flows led to the shocks seen in the VLBI observations, and inferred from the $\gamma$-ray detections.  

\begin{figure}
\begin{center}
\includegraphics[width=0.9\textwidth]{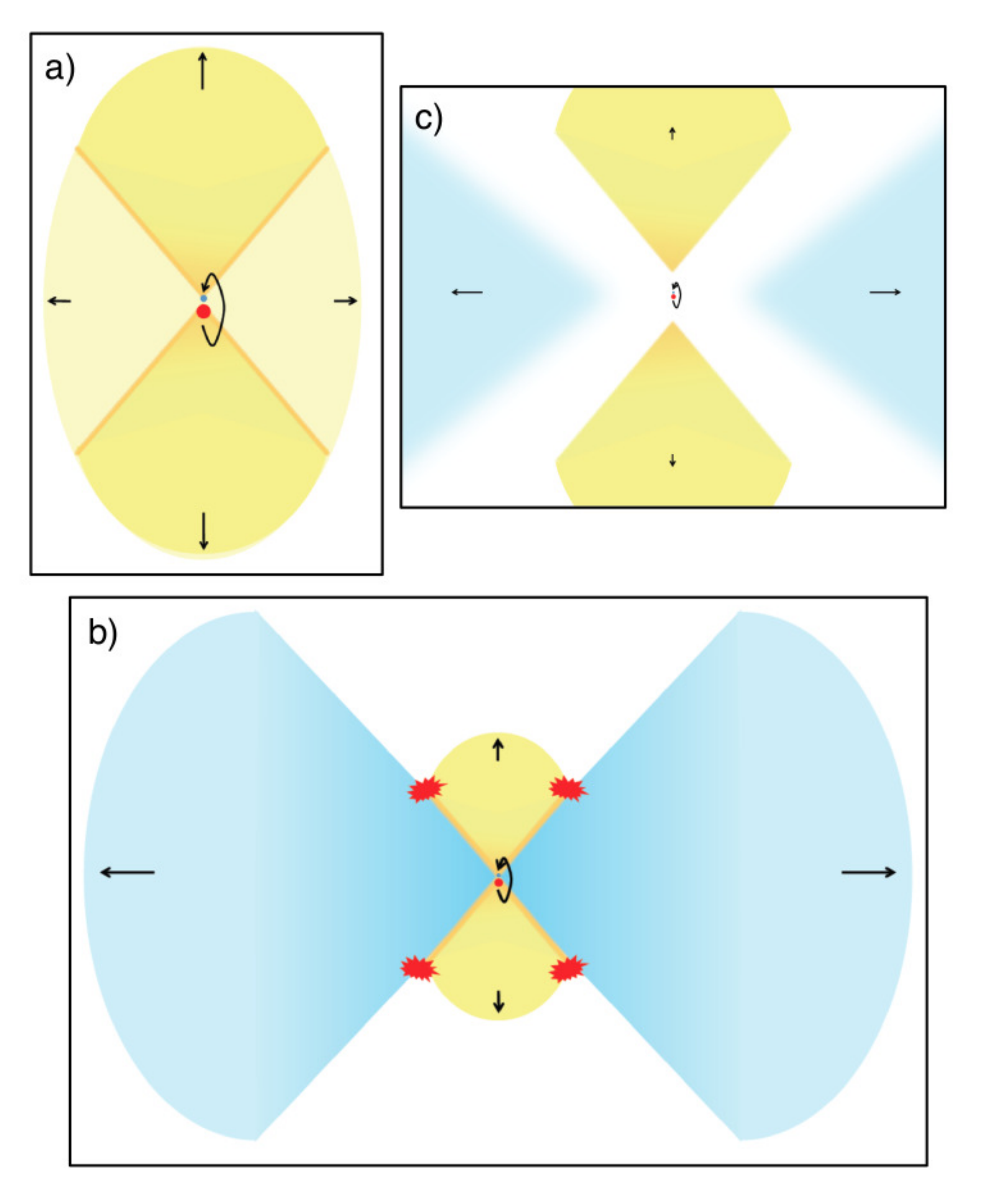}
\end{center}
\vspace{-0.6cm}
\caption{\textit{A cartoon picture to explain the evolution of V959 Mon. A) At early times, the ejecta interacts with the binary and a dense torus is formed (darker yellow, oriented vertically). B) Later, a fast wind (blue) shocks against the torus leading to synchrotron and $\gamma$-ray emission (red regions). This phase corresponds to the image of V959 Mon in Figure 1. C) At late times, the wind material is too diffuse to detect and the dense torus dominates the radio emission.  Adapted from Chomiuk et al. 2014.}\vspace{-12pt}}
\label{Mon_model}
\end{figure}

In our model (Figure~\ref{Mon_model}), the initial ejection of material from the nova is relatively slow (a few hundred km s$^{-1}$).  
This slow ejecta forms a common envelope around the binary and angular momentum is transferred such that a dense torus forms in the orbital plane (Porter et al. 1998).  Later, a fast wind develops and is confined by the torus to be ejected perpendicular to the orbital plane.
This fast flow dominated the VLA images during the first year of observing.  The slow equatorial flow in the torus remained optically thick much longer, and dominated the VLA images at late times.
The collision of the fast flow with the slower torus leads to shocks, creating a population of accelerated particles  (Chomiuk et al. 2014).  In V959 Mon, these accelerated particles were detected via radio synchrotron emission (Chomiuk et al. 2014).  There is new evidence that the shocks are also at least partially responsible for the optical emission.  Li et al. (2017) found a correlation between the optical and $\gamma$-ray light curves for V5856 Sgr that indicated the optical emission could be reprocessed shock emission.  

%\clearpage

In addition to uncovering the shocks, observations of V959 Mon in both the radio (Linford et al. 2015) and X-rays (Nelson et al., in prep.) suggest that the initial expansion stalled during the first month after eruption.  In the radio, the measured angular expansion rate of the bipolar material is inconsistent with expansion starting at the time of eruption.   Instead, the expanding material appears to have been ejected $\sim25$ days after the initial eruption. 
%(see Figure~\ref{Mon_expansion}). 
In the X-rays, there is an observed drop in column density towards the X-ray emitting region that is best described by a shell of material that was ejected $\sim30$ days after the nova was initially detected.

\section{Impact of the ngVLA: Sensitivity}
\label{sec:novae_ngvla_sensitivity}

The radio emission from classical novae provides information on the amount of mass ejected from the surface of the white dwarf.  Because thermal bremsstrahlung dominates the radio emission from the ejecta, radio observations trace the bulk of the ejecta more simply and more accurately than observations at other wavelengths (Seaquist \& Bode 2008).  For the first 100--200 days after eruption, the ejecta are optically thick at radio frequencies and the flux density is proportional to the electron temperature and the projected area.  This means that the sample of radio-detected novae is biased, since novae with small ejecta masses can only be detected if they are nearby (i.e., within a few kpc).  The increased sensitivity of the ngVLA will remove this bias by allowing the study of the full range of novae, with observations of even low-luminosity sources out to the Galactic Center.

\begin{figure}[t]
\begin{center}
\includegraphics[width=0.95\textwidth]{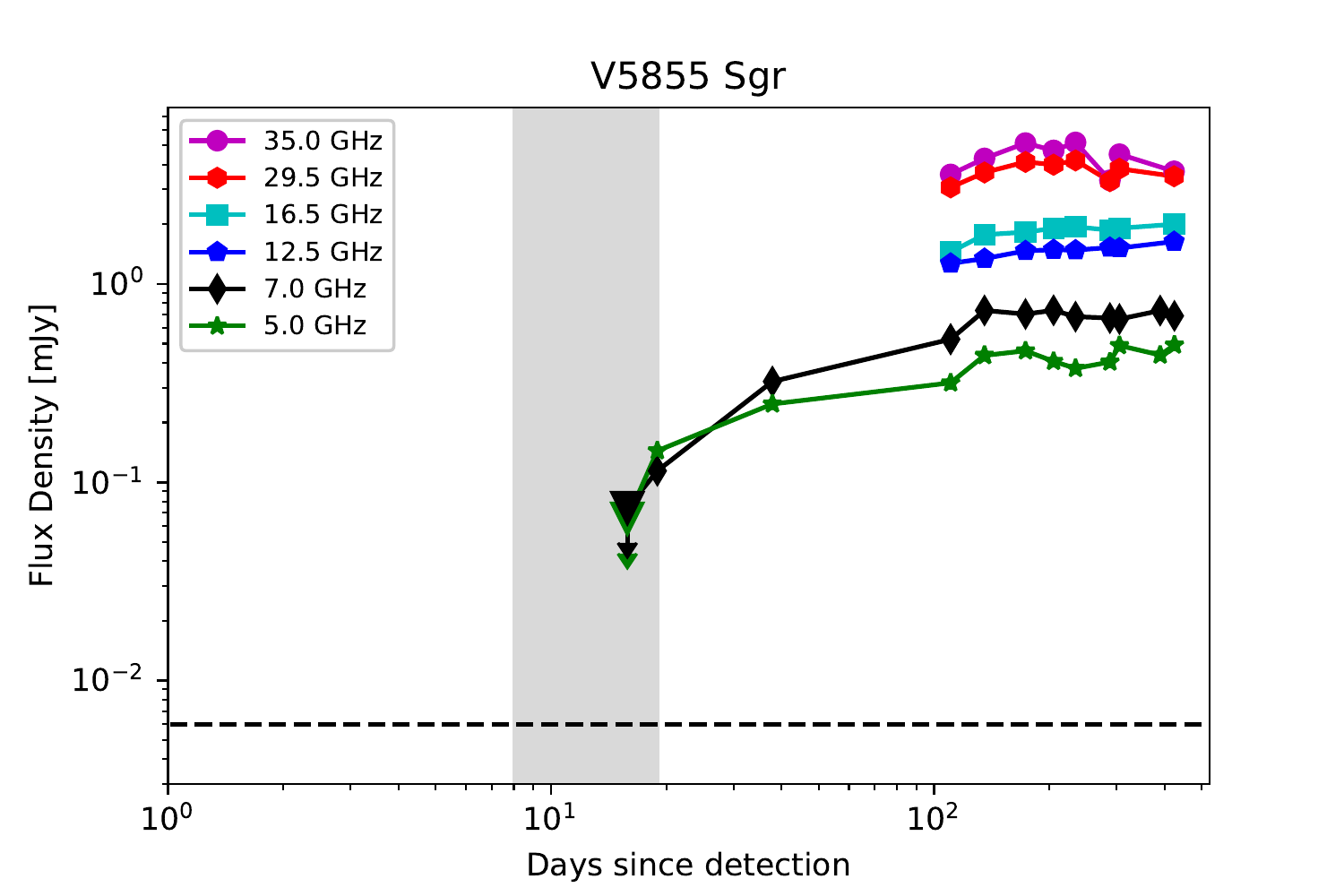}
\end{center}
\caption{\textit{The early radio light curve of V5855 Sgr.  Note the sudden transition from non-detection to detection at C-band (4-8 GHz).  Also, note that for the first detection, the 5 GHz flux density is higher than the 7 GHz flux density, which is opposite of what is expected for a source emitting via thermal bremsstrahlung.  The gray region indicates the time range while the nova was detected by the Fermi Gamma-ray Space Telescope.  The dashed line indicates the expected 3$\sigma$ RMS noise level of $\sim$6 $\mu$Jy/beam for 10 minute observations with 2 GHz of bandwidth centered between 4 and 12 GHz on the Next Generation VLA.}}
\label{TCP_lc}
\end{figure}

The recent discovery of $\gamma$-ray emission from classical novae renewed interest in understanding the causes and physics of shocks in classical novae.  One of the best tracers of shocks is synchrotron emission, which results in both steeply-falling radio spectra and high radio brightness temperatures ($\gtrsim 10^{5}$ K).  Spectral index measurements with the VLA have been moderately successful in identifying synchrotron emission early in a nova's evolution (e.g., Linford et al. 2017).  Another, more subtle indicator of shocks is the presence of strong radio emission at early times.  The source size may be inferred from its age and optical expansion velocities; if the radio flux density corresponds to brightness temperatures above $10^{5}$ K, there must be at least some contribution from synchrotron emission (e.g., Weston et al. 2016a,b).  This analysis rests on deep low-frequency (\textless7 GHz) observations early in the outburst, which have been rather sparse.  Even where there are such observations, one cannot tell how long the synchrotron (shock) emission persists, without more direct indicators, since the increasing size of the remnant makes lower brightness limits from the radio flux density less constraining as the source expands.  

The 2016 eruption of nova V5855 Sgr (TCP J18102829-2929590; CBET 4332) provided an intriguing test case.   
The first observation was a non-detection, while observations three days later showed strong emission. Moreover, the spectral index in the second epoch was characteristic of optically thin synchrotron emission, with $S_{\nu}\propto\nu^{-0.7}$.  This shows that 
early synchrotron emission can appear quite suddenly.  By the third observation, 19 days after the second epoch, the nova appeared to be emitting via optically thick thermal bremsstrahlung and no clear evidence for synchrotron emission remained.  Due to the sparse sampling of the light curve, it is not known how long the synchrotron emission persisted or if it showed any correlation with the $\gamma$-ray or X-ray emission.

With its high sensitivity, the ngVLA will be capable of uncovering synchrotron emission from less luminous and more distant novae. Shorter snapshot observations could be used to monitor novae at a high cadence (perhaps even daily) to look for the emergence and disappearance of synchrotron emission early in their evolution.  This will provide detailed information about the development and evolution of shocks in the nova ejecta, and 
allow direct comparison with high-cadence observations with X-ray and $\gamma$-ray instruments.

\section{Impact of the ngVLA: Imaging}
\label{sec:novae_ngvla_imaging}

High resolution ngVLA images will allow us to answer fundamental questions about novae. With a diversity of baseline lengths extending to 1000 km, every ngVLA observation of a nova has the potential to reveal structure like that imaged in Figure 1 --- simultaneously tracing both $\sim 10^{4}$ K thermal ejecta and compact synchrotron structure.

\noindent \textbf{What are the shapes of the expanding ejecta and how do they evolve?}  

It is still unclear whether nova ejecta are dominated by a spherical shell, a ring, or a bipolar structure (e.g., Slavin et al.~1995; Sokoloski et al.~2013). It is also unclear if nova explosions share a common ejecta morphology and shaping physics, or if their structures are shaped by a diversity of processes. ngVLA imaging will enable the search for evidence of a dense equatorial torus which could lead to the bipolar structure seen in V959 Mon (e.g., Ribeiro et al.~2009).  It is unknown exactly when this torus forms (or if it is already present prior to the eruption), so early images with higher angular resolution than the VLA allows are vital.  Understanding the evolving shape of a nova remnant is essential to interpreting observations at all wavelengths, from radio to $\gamma$-rays, and to testing the two-flow model. 

\begin{figure}[t]
\plottwo{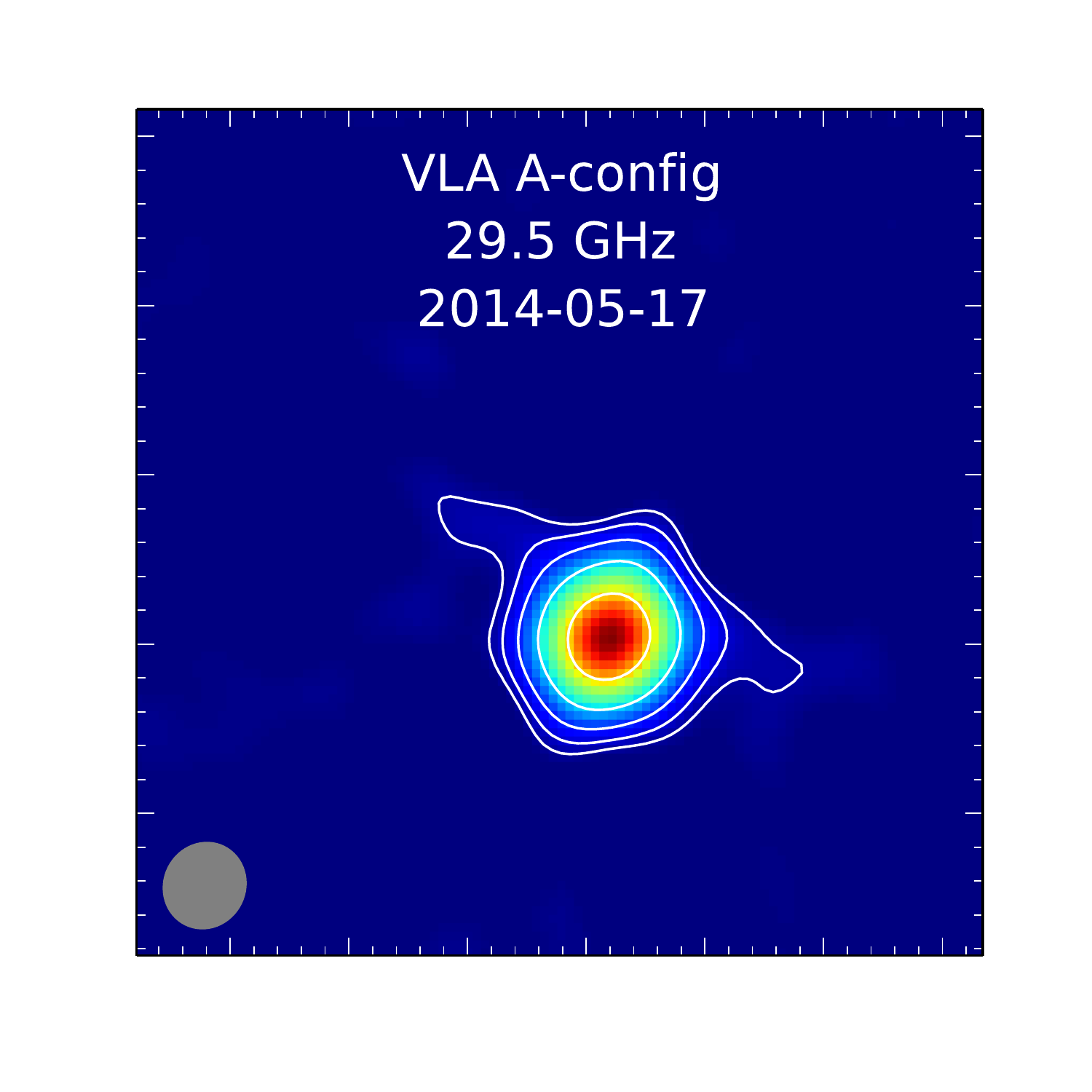}{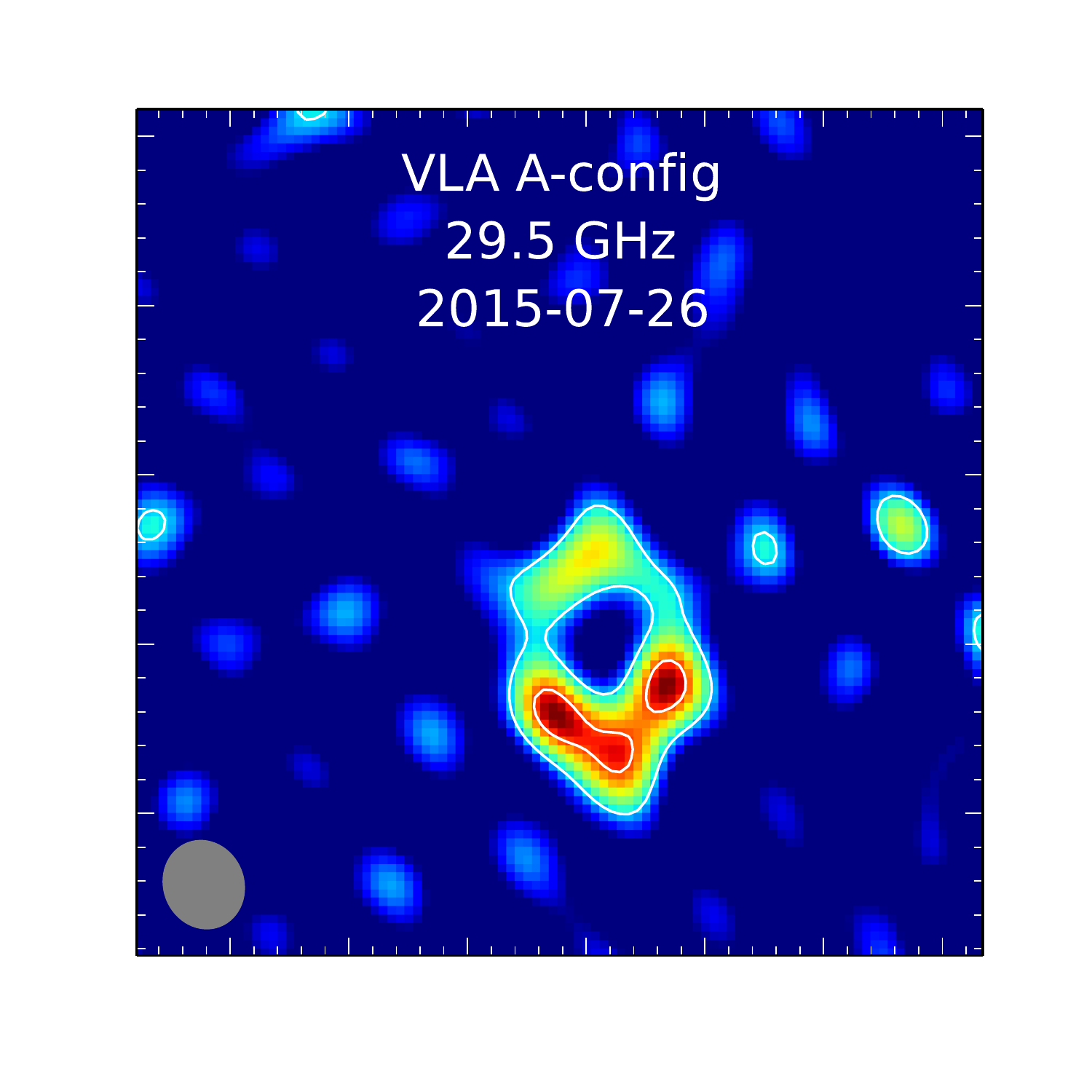}
\vspace{-0.9cm}
\caption{\textit{(Left) VLA A configuration image of V339 Del taken 2014-05-17.; (Right) VLA A configuration image of V339 Del taken 2015-07-26.  Note the appearance of a circular structure in the later image.  Due to the VLA configuration schedule, there was no way of resolving the the ejecta in between the A configurations. Over a year's worth of evolution must therefore be inferred from 2 snapshots.  The Next Generation VLA, with high angular resolution year-round, will for the first time allow imaging of the {\it entire} evolution of a nova.}\vspace{-12pt}}
\label{v339Del_2plot}
\end{figure}

The transition from the radio emission being dominated by the fast wind to being dominated by the slowly expanding torus has yet to be definitively captured, although Healy et al. (2017) reported a possible example in eMERLIN observations of V959 Mon.  In many novae the observations constrain only one of these phases.  In the Fermi-detected nova V339 Del, for example, there is evidence for the torus at late times, but no strong imaging evidence for the faster bipolar outflow (see Figure~\ref{v339Del_2plot}).  This is due in large part to the VLA only being in A configuration for roughly one quarter of the time.  Even in A configuration, the VLA only has sufficient angular resolution to resolve the ejecta in the nearest novae.  Having high sensitivity and high angular resolution of the ngVLA for every observation will capture the evolution from one phase to the other in detail,
in a single source, and show directly whether our hypothesized ejecta morphology is universal.

\noindent \textbf{Does the ejecta begin to expand immediately after eruption, or does the initial expansion stall?}

As discussed in Section~\ref{sec:novae_v959mon}, V959 Mon showed evidence for a stall in the expansion of the ejecta during the first month after eruption.  With current radio instruments, it is only possible to detect such a stall phase several months after the eruption and only for relatively nearby novae (within $\sim4$ kpc).  The ngVLA will have an order-of-magnitude better resolution and higher sensitivity than the VLA, and the resulting imaging should show both whether this such stalls are universal, and when and how the expansion resumes. The exact timing of this apparent stall can differentiate between a common envelope and a confinement phase (in which the nova ejecta are inhibited by external material present prior to the eruption).

\noindent \textbf{What is the energy budget of novae?} 

Combining ngVLA images of the expanding ejecta with velocity measurements from optical spectroscopy will allow for the application of expansion parallax techniques similar to those used for V959 Mon (Linford et al. 2015).  Knowing the distances to the novae will help determine whether only nearby novae are detected by \emph{Fermi} (e.g., Morris et al. 2017; Franckowiak et al. 2018), or if there is a large range in their $\gamma$-ray luminosities.  The $\gamma$-ray luminosities are also extremely valuable for understanding the energetics of these sources and determining whether the $\gamma$-ray production mechanism is leptonic or hadronic (e.g., Metzger et al. 2015; Vlasov et al. 2016).

\noindent \textbf{Where is the early synchrotron radiation?}

In addition to the expanding thermal ejecta, there has been evidence for early synchrotron emission in novae (e.g, Taylor et al. 1982; Weston et al. 2016a,b; Linford et al. 2017).  Presumably, this early non-thermal emission is related to the shocks that produce the $\gamma$-ray emission, but the relation is not clear (e.g., Vlasov et al. 2016).  For V959 Mon, For V959 Mon, observing both thermal and non-thermal components required a complicated campaign with multiple telescopes (VLA, EVN, VLBA, \& eMERLIN).  The higher sensitivity and longer baselines of the ngVLA will enable detection of synchrotron emission with higher confidence and earlier in the evolution of the nova than current capabilities allow.  The wide frequency range of the ngVLA will also allowing both the thermal and non-thermal components, simultaneously, and with a single telescope!

\section{Conclusions}

The ngVLA will be an amazing instrument for studying classical novae.  The increased sensitivity will enable detection of dimmer and more distant novae.  Fast wideband monitoring will allow probes of synchrotron emission from shocks at early times.  
The large range in baselines will enable simultaneous imaging of 
both the diffuse thermal ejecta and the regions of shock emission.  Finally, with high angular resolution at all times, every radio light curve data point will also be a frame in a movie tracing the evolution of the ejecta from shortly after the explosion to its final dispersal into the interstellar medium.  These movies will provide unprecedented opportunities to study these dynamic and enigmatic sources.

\acknowledgements We are grateful for the pioneering work of R.~M. Hjellming using the original VLA. %We are grateful to our collaborators C.~C. Cheung, T. Finzell, M.~I. Krauss, S. Lawrence, K-L. Li, B.~D. Metzger, A.~J. Mioduszewski, K. Mukai, T. Nelson, T.~J. O'Brien, V.~A.~R.~M. Ribeiro, N. Roy, J.~L. Sokoloski, J. Strader, A.~J. van der Horst, A. Vlasov, I. Vurm, C. Wendeln, J.~H.~S. Weston, J. Yang, and Y. Zheng for their contributions to the study of classical novae at radio wavelengths.   % Keep this text on the same line as the \verb"\acknowledgements" command because it makes things a lot easier.

%\bibliography{editor}  % For BibTex

% For non-BibTex:

\end{document}